\documentclass[11pt]{article}
\usepackage{jcappub} 
\usepackage{color}
\usepackage{latexsym}
\usepackage{graphicx}
\usepackage{subfigure}
\usepackage{hyperref}
\usepackage{amssymb}
\usepackage{bm}
\usepackage{natbib}
\usepackage{amsmath}
\title{Constraining warm dark matter with 21~cm line fluctuations due to minihalos}

\author[a]{Toyokazu Sekiguchi}
\author[b]{and Hiroyuki Tashiro}

\affiliation[a]{University of Helsinki and Helsinki Institute of Physics, 
P.O. Box 64, FI-00014, Helsinki, Finland}
\affiliation[b]{Physics Department, Arizona State University, Tempe AZ 85287, USA.}

\emailAdd{toyokazu.sekiguchi@helsinki.fi}

\abstract{
Warm dark matter (WDM) with mass $m_{\rm WDM}=\mathcal O(1)$~keV has long been
discussed as a promising solution for discrepancies between cosmic structures observed 
at small scales and predications of the concordance CDM model.
Though several cosmological observations such as the Lyman-alpha forest 
have already begun to constrain the range of $m_{\rm WDM}$,  
WDM is yet to be fully excluded as a solution for these so-called small-scale problems.
In this paper, we study 21~cm line fluctuations from minihalos in a WDM model
and evaluate constraints on $m_{\rm WDM}$ for future cosmological 21~cm surveys, such as SKA and FFTT.
We show that, since WDM with mass $m_{\rm WDM}\gtrsim10$~keV
decreases the abundance of minihalos by
suppressing the matter power spectrum on small scales via
free-streaming,
such WDM can significantly affect the resultant 21~cm line fluctuations from minihalos.
 We find that if the 21~cm signal from minihalos can be observed above $z\ge5$, 
 SKA and FFTT can give lower bounds $m_{\rm WDM}\gtrsim 24$~keV
and 31~keV, respectively, which are tighter than the current constraint.
These future 21~cm surveys might be able 
to rule out a WDM model as a solution of small-scale problems.
}

\begin{document}
\maketitle

\section{Introduction}

Cold dark matter (CDM) is a standard paradigm in modern cosmology~(for
a review, see~\cite{Frenk:2012ph}).
The CDM model predicts a hierarchical structure formation which is
strongly favored by observations of large scale structures.
However, some predictions of CDM model based on recent N-body simulations
seem to conflict with a number of observations on small scales~(see
\cite{Weinberg:2013aya} for a recent review).
One example is the so-called missing satellite problem: 
the number of observed satellite galaxies accompanying the Milky Way
is significantly smaller than the number seen in N-body simulations~\cite{Klypin:1999uc,Moore:1999nt}.
Another example is the so-called cusp-core problem: 
in simulations, dark matter halos have cuspy density profiles~\cite{Navarro:1996gj,Moore:1999gc},
that is, the density sharply increases at the center, while 
observations suggest the flat density profile around the center~\cite{Flores:1994gz,Moore:1994yx}.
Although these apparent discrepancies may originate from astrophysical
phenomena including baryon physics whose effects are not understood very
well, these might be suggesting
DM properties different from the standard CDM. 

Warm dark matter (WDM) with mass $m_{\rm WDM}\gtrsim 1$~keV 
has been paid particular attention as a potential solution of these discrepancies.
In a WDM model, since WDM particles remain ultra-relativistic after the
decoupling from the rest of the particles in the early universe, structure formation is suppressed
on small scales due to free-streaming of WDM particles~\cite{SommerLarsen:1999jx,Hogan:2000bv}.
As long as their mass is not very light, the WDM particles become highly non-relativistic 
in the late universe and end up to be similar to CDM particles.
The free-streaming length of a WDM particle can be approximately given by~\cite{SommerLarsen:1999jx}
\begin{equation}
\lambda_{\rm fs}=0.1\,{\rm Mpc}
\left(
\frac{\Omega_{\rm WDM}h^2}{0.11}
\right)^{1/3}
\left(
\frac{m_{\rm WDM}}{\rm keV}
\right)^{-4/3}.
\end{equation}
Within this scale, matter fluctuations are significantly suppressed.
This affects the formation of small dark matter halos and may
decrease the number of satellite dwarf galaxies~\cite{Schaeffer:1988,SommerLarsen:1999jx,
White:2000sy,Colin:2000dn}.
At the same moment, nonzero thermal velocity of WDM particles may
make the cusps of the density profile shallower than in the CDM model as
claimed in Ref.~\cite{Colin:2007bk}.

On the other hand, WDM can be constrained by various observations.
In particular, the suppression of density fluctuations on small scales due to WDM makes
the Lyman-alpha forest observation a powerful tool to constrain $m_{\rm
WDM}$~\cite{Narayanan:2000tp}.
The current observations of Lyman-alpha
provide the constraint~$m_{\rm WDM} \gtrsim 2.5$~keV at $2\sigma$ for 
a thermal relic WDM particle~\cite{Seljak:2006qw,Viel:2007mv,Viel:2013fqw}. 
In addition, 
high redshift observations of gamma-ray bursts can also place the
constraint ~$m_{\rm WDM} \gtrsim 1.6$~keV at 95\% CL~\cite{deSouza:2013wsa}.
The cosmic weak lensing has the potential to probe the small-scale
suppression induced by WDM. Future surveys like Euclid are expected to
put the constraint~$m_{\rm WDM} \gtrsim
2.5$~keV~\cite{Markovic:2010te,Smith:2011ev}.
There are also some works about the constraint on WDM mass
in the context of the cosmological reionization~\cite{Yoshida:2003rm,Yue:2012na}.
Some specific models of WDM can be constrained further via their particle properties.
For instance,
sterile neutrino is a well-known particle candidate for WDM.
In this case, a sterile neutrino decays into active ones, which leads to tighter constraints on its mass.
Observations of the X-ray background give an upper bound 
$m_{\rm WDM}< 4$~keV for sterile neutrino
mass~\cite{Boyarsky:2007ay}~(see also reference~\cite{Boyarsky:2008mt}).

In this paper, we focus on 21~cm line fluctuations induced by minihalos.
In Ref.~\cite{Loeb:2003ya},
21~cm line fluctuations from intergalactic medium (IGM) prior to the epoch of reionization
have been suggested as a promising probe of WDM.
Recently, Ref.~\cite{Sitwell:2013fpa} investigated the WDM effect on 21~cm signals
from IGM during the epoch of reionization, by taking into account that
WDM delays formation of halos of mass $M\gtrsim 10^7M_\odot$ needed for star-formation.
However minihalos of mass $M\lesssim10^7M_\odot$, 
which are not large enough to activate star formation, 
can also significantly contribute to 21 cm line signals around the epoch of
reionization~\cite{Iliev:2002gj,Furlanetto:2002ng}.
Future 21~cm surveys such as SKA\footnote{http://www.skatelescope.org} and FFTT~\cite{Tegmark:2008au} will
potentially observe these signals.
Since formations of small halos should depend
on the amplitude of matter fluctuations at corresponding scales, 
21~cm line fluctuations should also reflect matter fluctuations on these small scales, which
are difficult to be measured by CMB observations, large scale structure observations and
Lyman-alpha surveys.
Indeed, in Refs.~\cite{Takeuchi:2013hza, Sekiguchi:2013lma}, it is shown that 21~cm line fluctuations
from minihalos can be a powerful probe of isocurvature perturbations with blue-spectrum 
whose amplitude is large only at very small scales.
The 21~cm fluctuations due to minihalos can also probe the primordial
non-Gaussianity~\cite{Chongchitnan:2012we} and structure formations due
to cosmic strings~\cite{Tashiro:2013xra}. 
Although suppression in matter fluctuations 
is confined to very small scales $\lambda_{\rm fs}= \mathcal O(0.01)$Mpc
for a WDM model with rather large mass $m_{\rm WDM}\gtrsim 10$~keV, 
it may substantially affect 21~cm line fluctuations from minihalos 
through alternation of their abundance. 

In this paper, for simplicity we assume that WDM is fermion with internal degrees of freedom $g=2$
and has a thermal phase-space distribution. 
We also assume that dark matter consists only of WDM (or CDM) and do not consider mixed dark matter.
With this setup, the WDM model is parametrized by its mass $m_{\rm WDM}$ alone.
As rough estimation, we however remark that our results for thermal WDM 
would be translated into a variety of models of non-thermal WDM
by replacing the relation between $\lambda_{\rm fs}$ and $m_{\rm WDM}$.

This paper is organized as follows. In the next section, 
we discuss effects of WDM on the matter power spectrum and 
the halo mass function.
Density and temperature profiles of minihalos are also discussed here.
Then in Section~\ref{sec:21cm}, we study
21~cm line fluctuations in WDM model and constraints
on $m_{\rm WDM}$ expected for future 21~cm surveys.
We also examine impact of various effects of WDM on formation
and profile of halos due to its thermal velocity in Section~\ref{sec:profile}.
The final section is devoted to summary and discussion.

\section{Effects of WDM on mass function and inner profile of minihalos}
\label{sec:effects}

\subsection{Mass function}
\label{sec:massf}

\begin{figure}
  \begin{center}
    \hspace{-5mm}\scalebox{.9}{\includegraphics{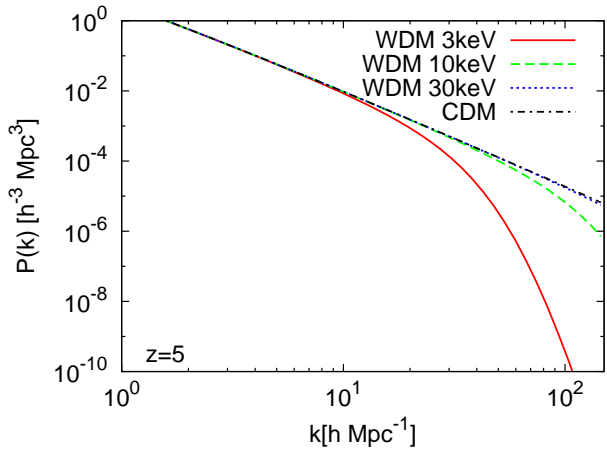}} 
  \end{center}
  \caption{Linear matter power spectra $P(k)$ in WDM and CDM models at redshift $z=5$.
  In the panel, $P(k)$ in WDM models with masses $m_{\rm WDM}=3$~(red solid), 10~(green dashed) 
  and 30~keV (blue dotted) as well as that in the CDM model (black dot-dashed) are plotted.
  Note that, since the suppression due to free-steaming of WDM given by Eq.~\eqref{eq:mpk}
  does not depend on redshift, the spectra $P_{\rm WDM}(k)$ as well as $P_{\rm CDM}(k)$ 
  at different redshifts differ only overall by growth factor ($\sim 1/(1+z)$ in the matter domination epoch).
  }
  \label{fig:mpk}
\end{figure}

In a WDM model, the linear matter power spectrum is suppressed on small
scales via free-streaming of WDM particles, 
compared to the one in the CDM model.  According to Ref.~\cite{Bode:2000gq}, 
the matter power spectrum in a WDM model
can be approximately given in terms of that in the CDM model as\footnote{We refer to 
Ref.~\cite{Boyanovsky:2010pw} for more detailed calculation of a linear transfer function in a WDM model.}
\begin{equation}
P_{\rm WDM}(k,z)=P_{\rm CDM}(k,z)
\left(1+(\alpha k)^{2\mu}\right)^{-10/\mu},
\label{eq:mpk}
\end{equation}
where $\mu=1.12$ and $\alpha$ is given by
\begin{equation}
\alpha=0.07\, {\rm Mpc}
\left(\frac{m_{\rm WDM}}{\rm keV}\right)^{-1.11}
\left(\frac{\Omega_{\rm WDM}h^2}{0.11}\right)^{0.11}.
\end{equation}
Fig.~\ref{fig:mpk} shows the linear matter power spectra in WDM models with $m_{\rm WDM}=3$, 10, 30~keV
in comparison with that in the CDM model. To compute $P_{\rm CDM}(k)$, the {\tt CAMB} 
code~\cite{Lewis:1999bs,Howlett:2012mh} is adopted here.
One can see that suppression from $P(k)$ in the CDM model occurs even on larger scales
as $m_{\rm WDM}$ becomes smaller, which represents the effect of WDM free-streaming.

Given a matter power spectrum or statistics of initial density perturbations, 
we can obtain the mass function of halos
based on the Press-Schechter formalism \cite{Press:1973iz}, 
\begin{equation}
\frac{dn}{d\ln M}=\bar\rho_m f(\nu)\frac{d\nu}{dM}, 
\end{equation}
where $\bar \rho_m$ is the mean energy density of matter and
$f(\nu)$ is the multiplicity function with significance of the critical overdensity $\delta_{\rm cr}\simeq1.67$ being
denoted as $\nu\equiv (\delta_{\rm cr}/\sigma_M)^2$.
Here, $\sigma_M$ is the rms of matter fluctuations in a sphere of radius $R$ satisfying 
$M=\frac{4\pi}3\bar\rho_m R^3$. 
Assuming a matter power spectrum $P(k,z)$, we can write $\sigma_M$ as
\begin{equation} 
\sigma_M^{2}= \int\frac{k^{2}dk}{2\pi^{2}}
W^{2}(kR)P(k),
\label{sigma8}
\end{equation}
where $W(kR)$ is a top-hat window function with radius $R$, 
\begin{equation} 
W(kR)\equiv \dfrac{3\left(\sin(kR)-kR\cos (kR) \right)}{\left( kR \right)^{3}}.
\end{equation}

Regarding the multiplicity function $f(\nu)$, we adopt the one proposed by Sheth and Tormen~\cite{Sheth:1999mn},
\begin{equation}
\nu f(\nu)=A\left(1+(q\nu)^{-a}\right)\left(\frac{q\nu}{2\pi}\right)^{1/2}\exp\left[-\frac{q\nu}2\right], 
\label{eq:occ}
\end{equation}
where $a=0.3$, $q=0.75$ and $A\simeq0.32$ are chosen to fit numerical results.

\begin{figure}
  \begin{center}
    \begin{tabular}{cc}
    \hspace{-5mm}\scalebox{.8}{\includegraphics{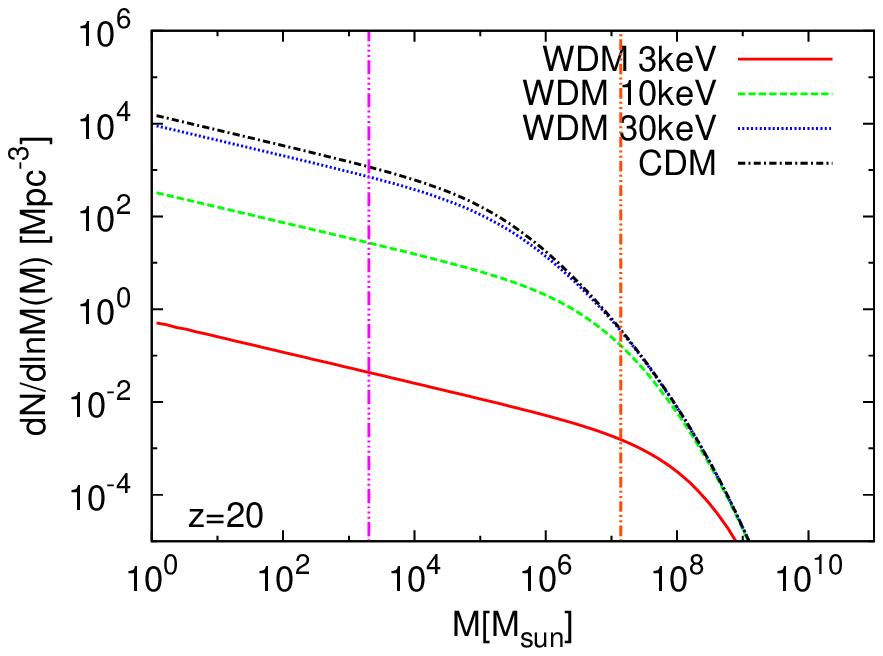}} &
    \hspace{-5mm}\scalebox{.8}{\includegraphics{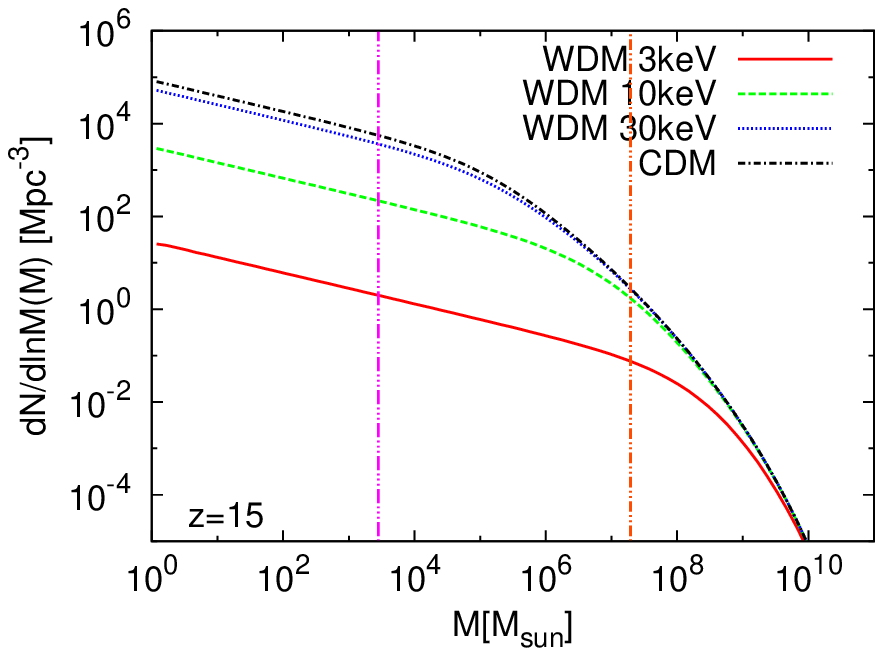}} \\
    \hspace{-5mm}\scalebox{.8}{\includegraphics{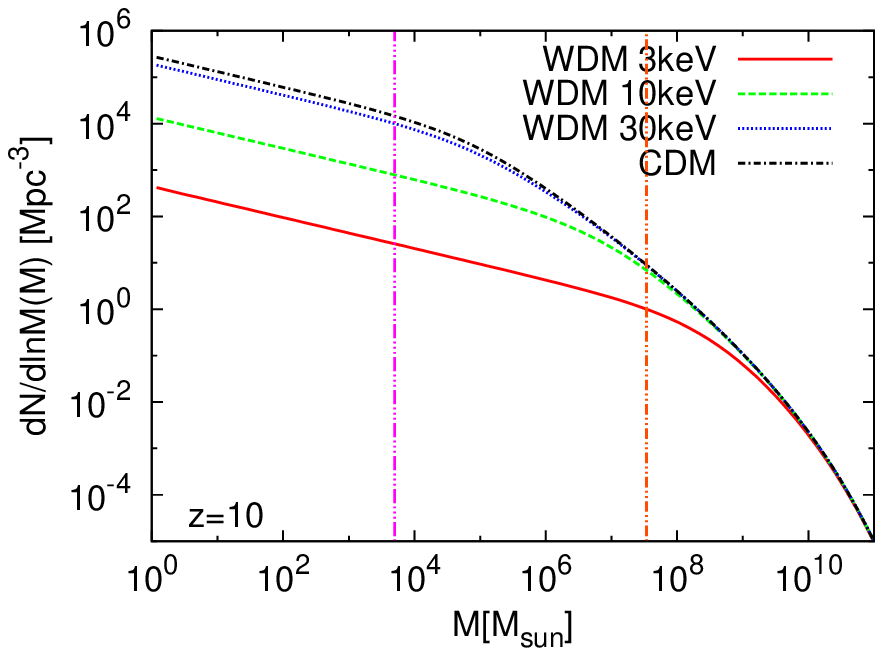}} &
    \hspace{-5mm}\scalebox{.8}{\includegraphics{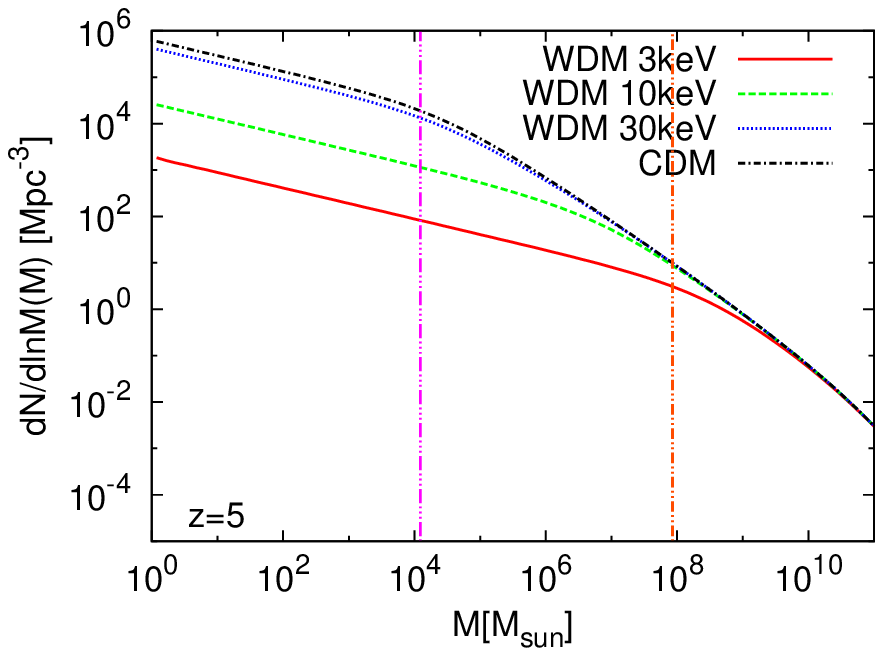}} 
    \end{tabular}
  \end{center}
  \caption{Mass functions are plotted. Adopted models of WDM and CDM are the same as in Fig.~\ref{fig:mpk}.
  Two vertical lines show $M_*=M(T_{\rm vir}=10^4{\rm K})$ (orange two-dot-chain) and the Jeans mass $M_J(z)$ (magenta three-dot-chain), 
  which respectively give the smallest and largest masses of minihalos~(see Eqs.~\eqref{eq:mj}-\eqref{eq:m*}).
  }
  \label{fig:massf}
\end{figure}

Mass functions in WDM and CDM models are plotted in Fig. \ref{fig:massf}.
We here also indicate the mass range contributing to 21~cm fluctuations from minihalos
(See Eqs.~\eqref{eq:m*}-\eqref{eq:mj}).
The figure explicitly shows that the abundance of halos are suppressed prominently at smaller masses.
We can also see that for $m_{\rm WDM}\gtrsim10$\, keV, while the abundance of halos which 
the star-formation $M\gtrsim10^7M_\odot$ is little affected, that of minihalos are still significantly suppressed.
This may indicate minihalos would be a more promising probe of WDM than halos which can host 
ionizing sources.

We should however note that the derivation of mass function above is
based on the hierarchical formation of
halos, which may break down in WDM models. This breakdown may lead that low mass halos form even further less.
This issue will be discussed in section~\ref{sec:profile}.

\subsection{Halo profile}

Our aim is to calculate the 21~cm signals from minihalos. The 21~cm
signal from a minihalo depends on the hydrogen density, temperature and
velocity dispersion. In this paper, we assume that a minihalo is modeled
as a nonsingular, truncated isothermal sphere ``TIS'' of dark matter and baryons in hydrostatic
equilibrium~\cite{Shapiro:1998zp}. 
In the TIS model, physical radius, gas temperature, dark matter
velocity dispersion and density profiles are respectively given as functions 
$r_t(M, z_{\rm coll})$, $T_K(l,M, z_{\rm coll})$, $\sigma_{\rm V}(l,M)$
and $\rho(l,M)$ where $l$ is the distance from the center of a minihalo
and $z_{\rm coll}$ is the redshift of the minihalo collapsing.

An overdense region with volume mass $M$ whose density contrast is
$\delta _M(z)$ at redshift $z$ collapses to a halo with mass $M$ at the
redshift given by
\begin{equation}
1+z_{\rm coll}=\frac{\delta_M(z)}{\delta_c}(1+z),
\end{equation}
where we use the fact that the growth factor
of matter density fluctuations is
proportional to $1/(1+z)$.

Since $\delta_M(z)$ fluctuates with the dispersion $\sigma_M$,
$z_{\rm coll}$ is also a random variable and
has a probability distribution. Then the mean $z_{\rm coll}$ of halos 
with $M$ which have already formed at redshift $z$ should be given by
\begin{eqnarray}
\langle 1+z_{\rm coll}\rangle (M,z)
&=&(1+z) \frac{\int^\infty_{\delta_c} d\delta_M
\frac{\delta_M}{\delta_c} \frac1{\sqrt{2\pi}\sigma_M(z)} 
\exp\left[-\frac{\delta_M^2}{2\sigma_M(z)^2}\right]}
{\int^\infty_{\delta_c} d\delta_M
\frac1{\sqrt{2\pi}\sigma_M(z)} 
\exp\left[-\frac{\delta_M^2}{2\sigma_M(z)^2}\right]} 
\notag\\
&=&(1+z) \left.
\frac{e^{-x^2}}{\sqrt{\pi}{x}\,{\rm erfc}(x)}
\right|_{x=\delta_c/\sqrt{2}\sigma_M(z)},
\label{eq:fluc_delta}
\end{eqnarray}
where ${\rm erfc}(x)=\frac2{\sqrt\pi}\int^\infty_xdt\,e^{-t^2}$ is the complementary 
error function. 
To understand Eq.~(\ref{eq:fluc_delta}) intuitively, let us consider two opposite limits. 
For $\delta_c/\sigma_M(z)\to \infty$, where only a tiny fraction
of density fluctuations $\delta_M$ has collapsed, one obtains $\langle 1+z_{\rm coll}\rangle\to1+z$, which
shows that 
most of existing such halos 
have just formed at $z$.
On the other hand, for $\delta_c/\sigma_M(z)\to 0$, where
a large fraction of $\delta_M$ has already collapsed, 
$\langle 1+z_{\rm coll}\rangle\to\sqrt{\frac2\pi}\frac{\sigma_M(z)}{\delta_c}(1+z)\gg1+z$, 
which shows that 
most of existing halos 
have already formed earlier.
We note that $1+z$ in the above argument should be in general replaced with 
inverse of the growth factor $1/D(z)$ in the $\Lambda$CDM cosmology.

Since, in a WDM model, $\sigma_M(z)$ is lower than in the CDM model,
density fluctuations $\delta_M$ take more time to grow and reach the 
critical overdensity $\delta_c$. Accordingly 
$\langle 1+z_{\rm coll}\rangle$ is smaller than in the CDM model, that is, the formation of halos effectively delays in a
WDM model.
This formation delay makes halos in WDM model less
concentrated.
As a result, the gas density and temperature inside decrease, 
while the sizes of halos broaden.
Along with the change in mass function, these changes in halo profiles
can also affect 21~cm line fluctuations from minihalos.

\section{21~cm fluctuations}
\label{sec:21cm}

Now we discuss 21~cm line fluctuations from minihalos whose mass is not large 
enough to host a galaxy. While minihalos are not luminous objects,
minihalos can make observable 21~cm line signals~\cite{Iliev:2002gj,Furlanetto:2002ng}.
To calculate the 21~cm signals from a minihalo, we follow the analysis in Refs.~\cite{Iliev:2002gj}. 
We also note that in this section, any parameter
dependences of quantities are significantly abbreviated.

We calculate the brightness temperature with rest-frame frequency $\nu'$
at a distance $r$ from the centre of a minihalo with mass $M$.
The brightness temperature is provided by
\begin{equation}
T_b(\nu',z,r,M)=T_{\rm CMB}(z)e^{-\tau_{\rm 21cm}}
+\int dR~T_{\rm s}(l)e^{-\tau_{\rm 21cm}(R)} \frac{d\tau_{\rm 21cm}(R)}{dR}.
\label{eq:Tb}
\end{equation}
Here $\nu_0$ is the frequency of 21~cm line emission in the rest frame, $\nu_0=1.4$ GHz,
$\tau_{\rm 21cm}(R)$ is the 21~cm optical depth along the photon path at $R$, which 
is given by\footnote{
Here we omitted the optical depth arising from the IGM.
Effects of the optical depth from the IGM in the brightness temperature (Eq.~\eqref{eq:Tb}) are significant only for 
very small minihalo masses, whose contributions in the total brightness temperature in Eqs.~\eqref{eq:dTbbar} and \eqref{eq:rms_dTb}
are irrelevant.
}
\begin{equation}
\tau_{\rm 21cm}(R)=\frac{2c^2A_{10}T_*}{32\pi\nu_0^2}
\int^R_{-\infty} \frac{n_{\rm HI}(l')\phi(\nu',l')}{T_{\rm s}(l')}dR', 
\label{eq:tau_halo}
\end{equation}
where $l'=\sqrt{R^{\prime2}+r^2}$, $A_{10}=2.85\times 10^{-15}$ s$^{-1}$ and $k_BT_*=h\nu_0=5.9\times10^{-6}$ 
eV are respectively the spontaneous decay rate and emitted energy of the 21~cm hyperfine transition, and
$n_{\rm HI}$ is the number density of neutral hydrogen atoms. 
As the line profile $\phi(\nu',l')$, we adopt the thermal
Doppler-broadening model, i.e. 
$\phi(\nu',l')=\left(\sqrt{\pi}\Delta \nu(l') \right)^{-1} \exp[-(\frac{\nu'-\nu_0}{\Delta \nu(l')})^2]$ with
$\Delta \nu=(\nu_0/c)\sqrt{2k_BT_K(l')/m_H}$, where $m_H$ is the hydrogen mass.
The optical depth $\tau_{\rm 21cm}$ in Eq.~\eqref{eq:Tb} is given as $\tau_{\rm 21cm}=\tau_{\rm 21cm}(R\to\infty)$.
In 21~cm observations,
observation signals are measured in terms of the differential brightness temperature
\begin{equation}
\delta T_b(\nu;z,r,M)=T_b(\nu',z,r,M)/(1+z) -T_{\rm CMB}(0).
\label{eq:dTb}
\end{equation}
Therefore, the differential line-integrated flux from this halo is given by~\cite{Iliev:2002gj}
\begin{equation}
\delta  F = \int d\nu'~ 2 \nu'^2 k_B \langle \delta T_b \rangle(\nu';z,M)
 A(M,z),
\end{equation}
where $\langle \delta T_b \rangle(\nu';z,M)$ is the mean surface brightness temperature for a halo with mass $M$ at
$z$ is obtained from
\begin{equation}
\langle \delta T_b \rangle (\nu;z,M)
 = \frac{1}{A(M)} \int dr~ 2\pi r \delta T_b(\nu;z,r,M) ,
\end{equation}
where $A(M, z)$ is the geometric cross-section of a halo with mass $M$ at $z$,
$A=\pi r_t^2$.

Let's consider an observation with a finite bandwidth $\Delta \nu$ and
beam width $\Delta \theta$ .
The mean differential flux per unit frequency is written as
\begin{equation}
 \frac{d \delta F}{d \nu} =
\frac{\Delta z(\Delta\Omega)_{\rm beam}}{\Delta\nu}
\frac{d^2V(z)}{dz\,d\Omega}
\int^{M_{\rm max}(z)}_{M_{\rm min}(z)} dM
\delta F
\frac{dn}{dM}(M,z),
\label{eq:dif_flux}
\end{equation}
where $(\Delta\Omega)_{\rm beam}=\pi(\Delta\theta/2)^2$,
and $\Delta\nu/\Delta z=\nu_0/(1+z)^2$. 
In Eq.~(\ref{eq:dif_flux}), $M_{\rm min}(z)$ and 
$M_{\rm max}(z)$
are respectively the minimum and maximum masses of minihalos.
We assume that $M_{\rm min}(z)$ corresponds to the Jeans mass
\begin{equation}
M_{\rm J}(z)=5.73\times 10^3\left(\frac{\Omega_mh^2}{0.15}\right)^{-1}
\left(\frac{\Omega_bh^2}{0.022}\right)^{-3/5}
\left(\frac{1+z}{10}\right)^{3/2}M_\odot.
\label{eq:mj}
\end{equation}

Halos with mass larger than $M_{\rm max}(z)$ can host stars or galaxies and
most of the hydrogen in the halos is ionized. Therefore, such halos cannot
contribute to 21~cm line signals.
We set $M_{\rm max}(z)$ to the virial mass $M_*(z)$ with a virial
temperature $10^4$K which corresponds to the critical temperature for
hydrogen atomic cooling.
According to Ref.~\cite{Iliev:2002gj}, 
$M_*(z)$ can be approximately given as
\begin{equation}
M_*(z)=3.95\times 10^7\left(\frac{\Omega_mh^2}{0.15}\right)^{-1}
\left(\frac{1+z}{10}\right)^{-3/2}
M_\odot.
\label{eq:m*}
\end{equation}

Defining the beam-averaged ``effective'' differential antenna temperature $\overline{\delta T}_b$ by
$d \delta F /d \nu=2\nu^2k_B
\overline{\delta T_b}(\Delta\Omega)_{\rm beam}$, we finally obtain
\begin{equation}
\overline{ \delta T_b}(\nu)\approx
2\pi c\frac{(1+z)^4}{H(z) \nu_0}\int^{M_{\rm max}(z)}_{M_{\rm min}(z)}
dM~\frac{dn}{dM}(M,z) \Delta\nu_{\rm eff}(z)  A(M,z)\langle \delta T_b \rangle(z,M).
\label{eq:dTbbar}
\end{equation}

In future observations, important observable values are the fluctuations
of $\overline{\delta T_b}(\Delta\Omega)_{\rm beam}$. Since a minihalo is
a biased tracer of density fluctuations, the number density contrast of
minihalos are given by 
\begin{equation}
 \delta _N (M) = b(M) \delta ,
\end{equation}
where $\delta_N(M)$ is the number density contrast of halos with mass
$M$ and $b(M)$ is bias. Given the multiplicity function in Eq.~\eqref{eq:occ}, 
we adopt the following $b(M)$~\cite{Sheth:1999mn}: 
\begin{equation}
b(M) = 1+\frac{q\nu-1}{\delta_{\rm cr}} + \frac{2a}{\delta_{\rm cr}[1+(q\nu)^p]}.
\end{equation}
This clustering causes fluctuations in $\overline{ \delta T_b}(\nu)$.
Because the rms density fluctuations in a beam can be written as
\begin{equation}
\sigma_p^2(\nu, \Delta \nu, \Delta \theta) =
\int \frac{d^3k}{(2\pi)^3}W(\vec k; \nu,\Delta\nu, \Delta \theta)^2P(k),
\end{equation}
where $W$ is a pencil beam window function at frequency $\nu$ with band width $\Delta \nu$ and beam width $\Delta\theta$
(see e.g. Ref.~\cite{Dodelson:2003ft}), 
the rms fluctuations of the halo number density with mass $M$ in a beam is provided by
\begin{equation}
 \sigma_N(M) = b(M) \sigma_p(\bar \nu, \Delta \nu, \Delta \theta).
\end{equation}
Accordingly, we can obtain the rms fluctuations of $\delta T_b(\nu)$ 
from
\begin{equation}
\sigma_{\delta T_b}=\langle \delta T_b(\nu)^2 \rangle^{1/2}
\approx \overline{\delta T_b}(\nu)  \beta(z) \sigma_p(\bar \nu, \Delta \nu, \Delta \theta),
\label{eq:rms_dTb}
\end{equation}
where $\beta(z)$ is the effective bias of the minihalos weighted by their 21~cm line fluxes, defined as
\begin{equation}
\beta(z)\equiv \frac{\int^{M_{\rm max}(z)}_{M_{\rm min}(z)}  dM \frac{dn}{dM}(M,z) \mathcal F(z,M) b(M,z)}
{\int^{M_{\rm max}(z)}_{M_{\rm min}(z)}  dM \frac{dn}{dM}(M,z) \mathcal F(z,M)},
\end{equation}
where $\mathcal F(z,M)\equiv\int d^2b~\delta T_b(z,b,M)\propto T_br_t^2 \sigma_V$~\cite{Chongchitnan:2012we} is the flux from a minihalo.

\begin{table}
  \begin{center}
    \begin{tabular}{lcc}
      \hline\hline
      & SKA & FFTT \\
      \hline
      antenna collecting area $A$ [m$^2$]& $10^5$ & $10^7$ \\
      bandwidth $\Delta \nu$ [MHz] & \multicolumn{2}{c}{$1$} \\
      beam width $\Delta \theta$ [arcmin] & \multicolumn{2}{c}{$9$} \\
      integration time $t$ [hour] & \multicolumn{2}{c}{$10^3$}  \\
      \hline\hline
    \end{tabular}
  \end{center}
  \caption{Survey parameters for the SKA and FFTT surveys}
  \label{tbl:survey}
\end{table}
\begin{figure}
  \begin{center}
  \scalebox{.9}{\includegraphics{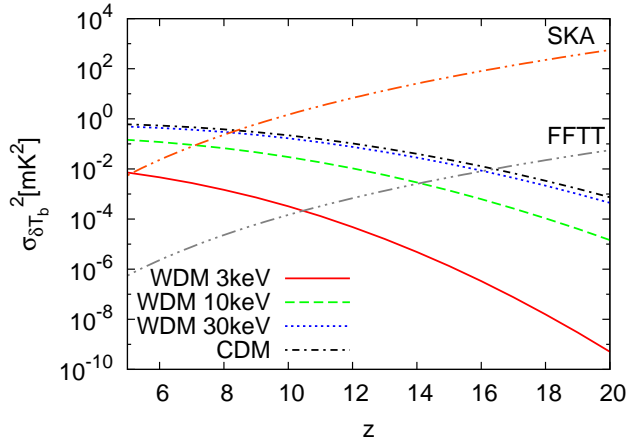}}
  \end{center}
  \caption{Rms of 21~cm line fluctuations $\sigma_{\delta T_b}^2$ are plotted as functions of redshift 
  in WDM models with $m_{\rm WDM}=3$ (red solid), 10 (green dashed), 30 (blue dotted)  
  as well as in the CDM model (black dot-dashed). As reference, expected noise levels of SKA (orange two-dot-chain) and 
  FFTT (magenta three-dot-chain) are also plotted.
  }
  \label{fig:rms}
\end{figure}
\begin{figure}
  \begin{center}
    \hspace{-5mm}\scalebox{.8}{\includegraphics{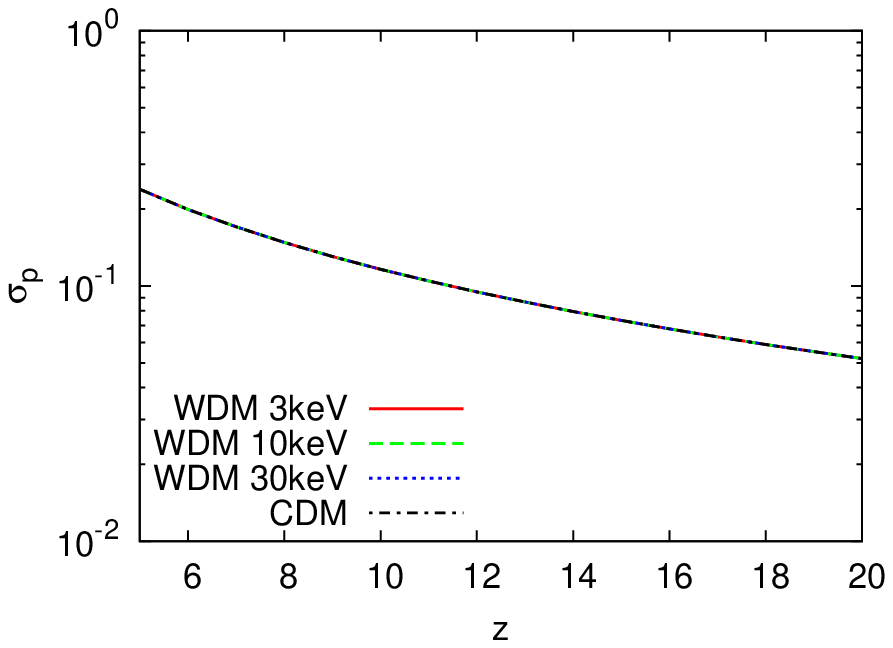}} \\
    \hspace{-5mm}\scalebox{.8}{\includegraphics{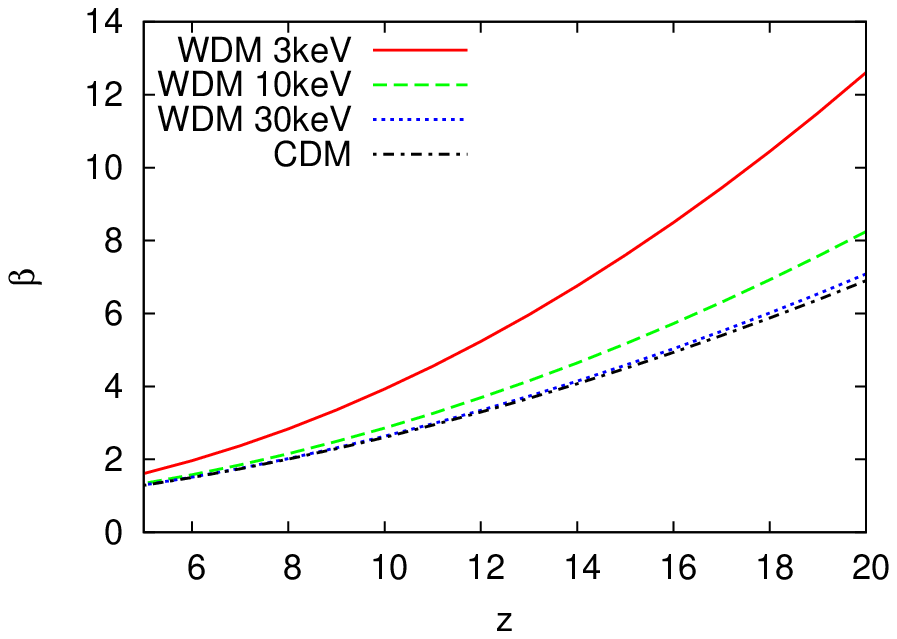}} \\
    \hspace{-5mm}\scalebox{.8}{\includegraphics{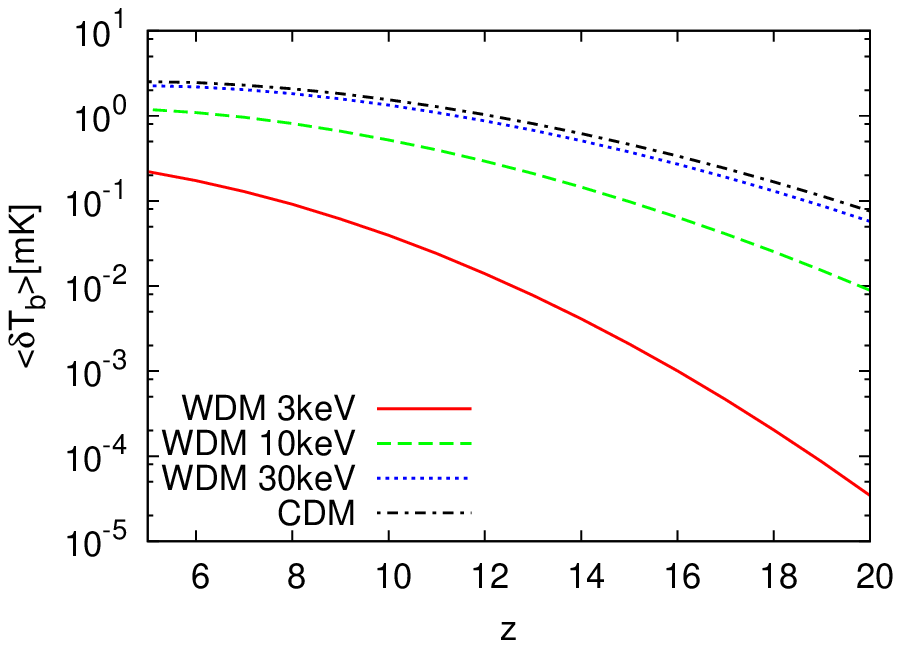}}
  \end{center}
  \caption{The three components in Eq.~\eqref{eq:rms_dTb} are plotted separately. 
  In order from top to bottom, the smoothed matter fluctuations $\sigma_p$ for the resolution of SKA,
  the effective bias $\beta$ and average differential brightness temperature $\overline{\delta T}_b$ are plotted.
  In each panel, adopted models of WDM and CDM are the same as in
  Fig.~\ref{fig:rms}.
  We note that all lines in the top panel overlap with one another.
  }
  \label{fig:comp}
\end{figure}

In Fig.~\ref{fig:rms}, we plot the 21~cm line fluctuations $\sigma_{\delta T_b}^2$ as function of redshift,
adopting the resolution $\Delta \theta$ and the bandwidth $\Delta \nu$ in Table~\ref{tbl:survey}. 
One can see that the fluctuations are suppressed in WDM models compared to those in the CDM model.
The degree of suppression depends on $m_{\rm WDM}$. For $m_{\rm WDM}=30$~keV, the suppression 
is almost negligible, and the resultant 21 cm fluctuations are almost indistinguishable from those in the CDM model. 
On the other hand, the suppression becomes larger for smaller $m_{\rm WDM}$."
This result is largely as is expected straightforwardly from the observations of mass function
in the previous section, that is, the abundance of minihalos is suppressed in WDM models.

To understand the origin of WDM effects on 21~cm line fluctuations,
we plot three components of Eq.~\eqref{eq:rms_dTb} separately in Fig.~\ref{fig:comp}.
We can see that the smoothed matter fluctuations $\sigma_p$ 
hardly changes over the range of $m_{\rm WDM}$ considered here.
This is because the comoving size of a {\it boxel} specified by $\Delta\theta$ and $\Delta\nu$
is much larger than the
free-streaming scale of WDM with mass $m_{\rm WDM}>1$~keV.
On the other hand, the effective bias $\beta$ increases for smaller $m_{\rm WDM}$.
This effect works on 21~cm line fluctuations in an opposite manner compared with the effect of WDM through the halo mass function.
However, the increase in effective bias is at most several tens of percents and cannot rival the 
influence from the mass function. Finally, we can see that the average temperature $\overline{\delta T_b}$ 
depends on $m_{\rm WDM}$ much more significantly compared to above two components. 
This dependency of $\overline{\delta T_b}$ on $m_{\rm WDM}$ almost completely accounts for
effects of WDM on $\sigma_{\delta T_b}^2$. 
We again note here that $\overline{\delta T_b}$ and hence $\sigma_{\delta T_b}^2$ 
for $m_{\rm WDM}>30$~keV are almost indistinguishable from those in the CDM model.

Now let us briefly discuss expected constraints on WDM models from future observations of 21~cm line fluctuations.
According Ref. \cite{Furlanetto:2006jb}, 
the noise level of the 21~cm line survey can be approximately given by
\begin{equation}
\sigma_{dT_b}^{\rm (noise)}=20\mbox{mK}\left(\frac{A}{10^4 \mbox{m}^2}\right)^{-1}
\left(\frac{\Delta \theta}{10^\prime}\right)^{-2}
\left(\frac{1+z}{10}\right)^{4.6}
\left(\frac{\Delta \nu}{\mbox{MHz}}\frac{t}{100\mbox{h}}\right)^{1/2}.
\end{equation}
In Fig.~\ref{fig:rms}, ${\sigma_{dT_b}^{\rm (noise)}}^2$ for SKA and FFTT,
whose survey parameters are summarized in Table~\ref{tbl:survey}, are plotted along with cosmological 
signals. One can see that these future observations can distinguish
a WDM model with $m_{\rm WDM}=10$~keV, which is allowed by current observations, from the CDM one.
On the other hand, 
because the difference between the rms of the 21 cm line fluctuations in the CDM model and WDM ones with $m_{\rm WDM}\gtrsim 30$~keV
becomes small, it is difficult to discriminate between the CDM model and such massive WDM ones.

\begin{figure}
  \begin{center}
  \scalebox{.9}{\includegraphics{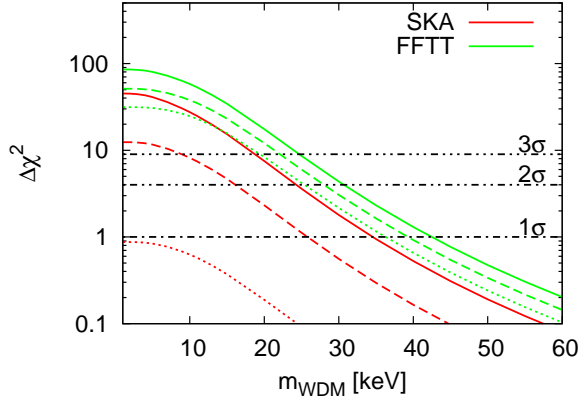}}
  \end{center}
  \caption{Forecast for constraints on WDM models from observations of 21~cm line fluctuations. 
  $\Delta \chi^2$ expected for SKA (red) and FFTT (green) surveys are plotted as functions of $m_{\rm WDM}$.
  For each observation, we take three different values of the minimum observable redshift: 
  $z_{\rm min}=5$ (solid), 7 (dashed) and 9 (dotted) (See the text for more details).
  Black horizontal lines represent deviation from CDM model at 1 (dot-dashed), 2 (two-dot-chain), and 3\,$\sigma$ (three-dot-chain) levels.
  }
  \label{fig:dchi2}
\end{figure}

To be more quantitative, we compute $\Delta \chi^2$ in WDM models by taking the CDM model as fiducial one.
Here we simply assume that each band in the observed frequency range is 
independent and has the same survey parameters as in Table~\ref{tbl:survey}.
Then $\Delta \chi^2$ can be given as
\begin{equation}
\Delta \chi^2(m_{\rm WDM})=\frac12
\sum_{\rm bands} \left[\frac{{\sigma_{\delta T_b}}^2(m_{\rm WDM})-{\sigma^{\rm (fid)}_{\delta T_b}}^2}
{{\sigma^{\rm (fid)}_{\delta T_b}}^2+{\sigma^{\rm (noise)}_{\delta T_b}}^2}\right]^2, 
\label{eq:s2n}
\end{equation}
where $\sigma^{\rm (fid)}_{\delta T_b}$ denotes the rms 21~cm line fluctuations 
in the CDM model.
In Fig.~\ref{fig:dchi2}, we plot $\Delta \chi^2$ of WDM models expected for SKA and FFTT.
As can be read from Fig.~\ref{fig:rms}, since the signal to noise ratio becomes more and more prominent at lower redshifts,
$\Delta \chi^2$ is dependent on the minimum redshift $z_{\rm min}$ above which 21~cm line fluctuations from minihalos can be observable.
In order for this dependency to be captured, in Fig.~\ref{fig:dchi2}, 
we take three different values of $z_{\rm min}=5$, 7 and 9, while the maximum observable 
is fixed to 20.
If the 21~cm signal from minihalos can be observable above $z_{\rm min}=5$ (7), 
SKA and FFTT can respectively  give lower limits on $m_{\rm WDM}$ as 24 (16)~keV and 31 (28)~keV at 2$\sigma$ level.
On the other hand, if the signal from minihalos can be observable only above $z_{\rm min}=9$, SKA would not be 
sensitive enough to give a constraint on $m_{\rm WDM}$, while FFTT can still give a lower limit $m_{\rm WDM}\ge$25~keV (2$\sigma$).

\section{Impact of effects on formation and profile of halos}
\label{sec:profile}

So far we have taken into account effects of WDM only through the
transfer function as shown in Eq.~(\ref{eq:mpk}), 
which are brought about when WDM is relativistic.
However, WDM can provide additional effects on the formation and structure of small halos.
Such effects basically enhance the difference in 21~cm line fluctuations between WDM and CDM models, 
and hence the constraints on $m_{\rm WDM}$ in the previous section
could be modified by taking into account these effects.
In this section, we include such effects of WDM on formation and profile of halos
in a phenomenological manner and examine whether these effects are likely to 
change the constraints furthermore.

The number density of halos, Eq.~(\ref{eq:occ}), is based on the hierarchical formation of halos.
However, since fluctuations within free-streaming scale are suppressed
in a WDM model as shown in Eq.~(\ref{eq:mpk}),
this hierarchical picture should break down around the corresponding scale.
Thus, the abundance of halos would be significantly altered below the so-called 
free-streaming mass $M_{\rm fs}=\frac{4\pi}3\bar\rho_m(\lambda_{\rm fs}/2)^3$.
 Actually, a number of studies with numerical simulations have claimed that
there should be a cut-off in the mass function around $M_{\rm fs}$
and the abundance of halos below this cut-off should be significantly suppressed
\cite{Zavala:2009ms,Polisensky:2010rw,
Benson:2012su,Schneider:2013ria,Angulo:2013sza}\footnote{This issue seems not completely settled yet.
See also Ref.~\cite{Wang:2007he}.}.
However Eq.~(\ref{eq:occ}) predicts substantial abundance of halos.
Following Ref.~\cite{Smith:2011ev}, we take into account this suppression by introducing a 
cut-off in the mass function, 
\begin{equation}
\frac{d\tilde n}{dM}=
\frac12
\left[1+{\rm erf}\left\{
\frac{\log_{10}(M/M_{\rm fs})}
{\sigma_{\log_{10}M}}
\right\}\right]
\frac{dn}{dM}.
\end{equation}

\begin{figure}
  \begin{center}
    \begin{tabular}{cc}
    \hspace{-5mm}\scalebox{.8}{\includegraphics{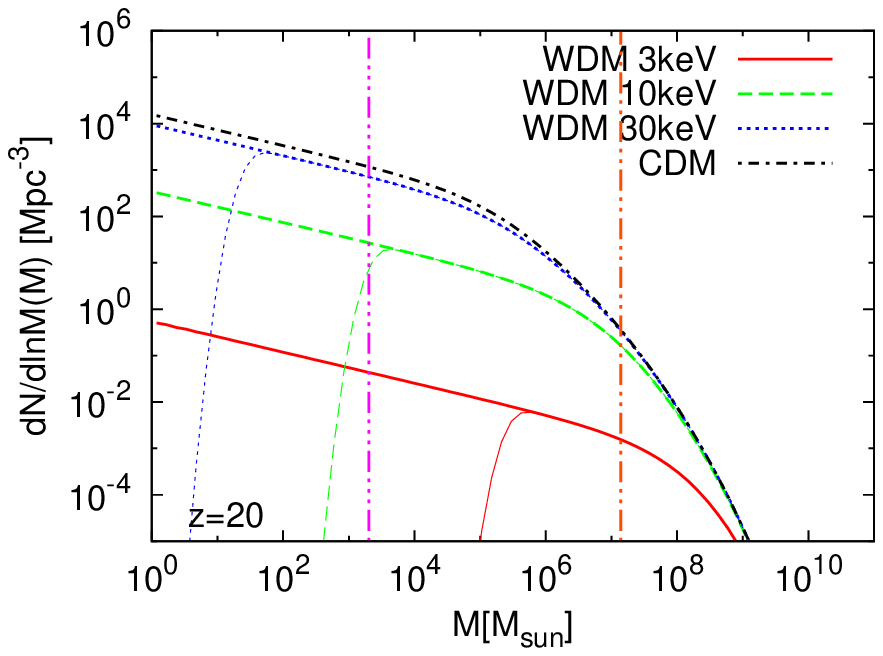}} &
    \hspace{-5mm}\scalebox{.8}{\includegraphics{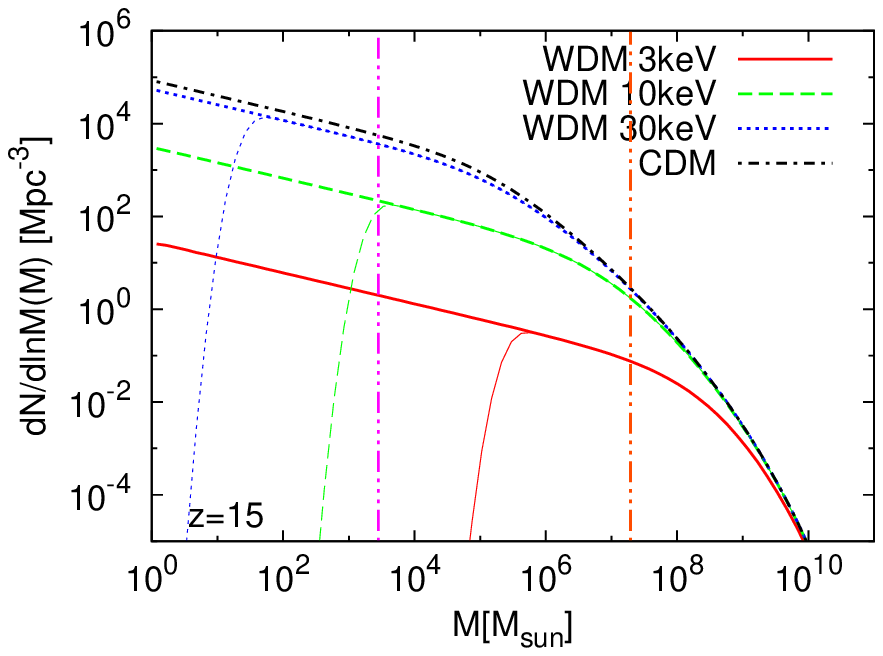}} \\
    \hspace{-5mm}\scalebox{.8}{\includegraphics{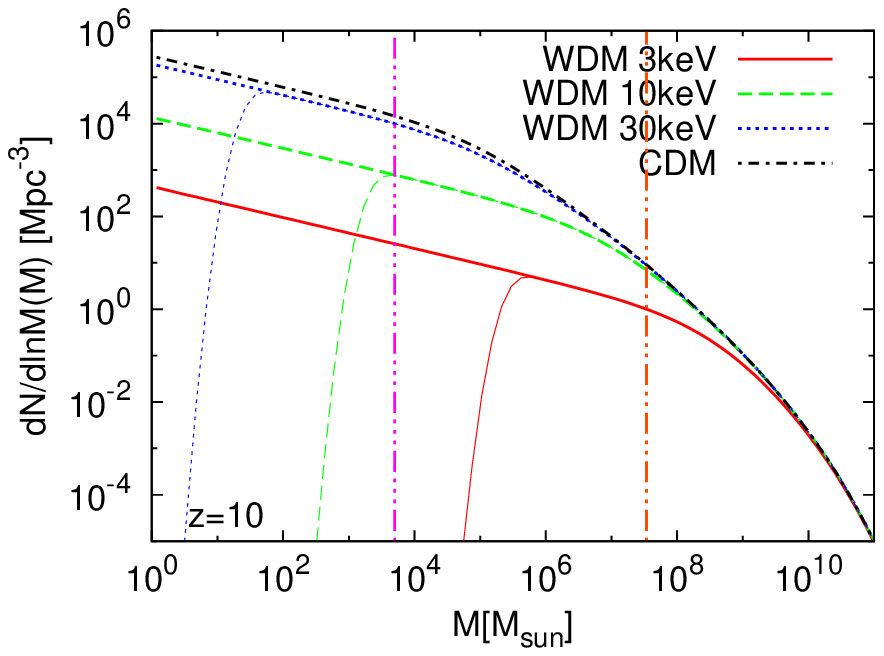}} &
    \hspace{-5mm}\scalebox{.8}{\includegraphics{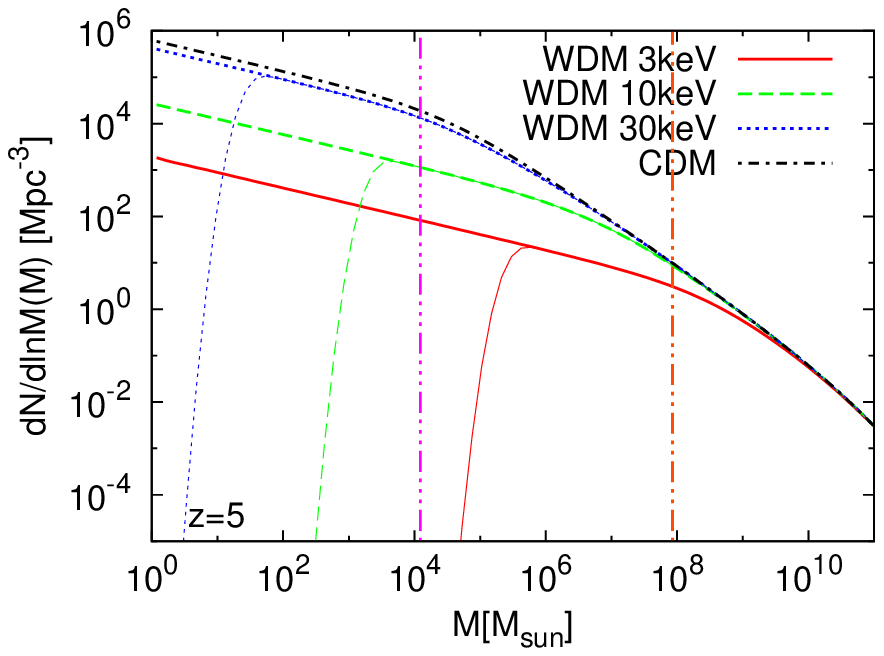}} 
    \end{tabular}
  \end{center}
  \caption{Comparison of mass function in WDM models with (thin) and without (thick) cut-off at $M_{\rm fs}$. 
  Models of WDM and CDM are the same as in Fig.~\ref{fig:massf}.
  }
  \label{fig:massf_wcutoff}
\end{figure}

Fig.~\ref{fig:massf_wcutoff} shows mass functions in WDM models with and without the cut-off.
Depending on the WDM mass, the cut-off scales differs. In the cases of WDM
mass larger than $10$~keV, the cut-off mass becomes smaller than the
Jeans mass which is  the minimum mass of minihalos in this paper.
As will be confirmed later, this indicates that presence of cut-off would be 
relevant for 21~cm fluctuations from minihalos only when $m_{\rm WDM}\gtrsim10$~keV.

Due to the thermal velocity of WDM, DM in halos has large 
phase space volume in a WDM model~\cite{Smith:2011ev}.
This makes less DM particles confined at the central region
of a halo and hence the density profile ends up to be smoothed.
To take into account this effect,
we follow the approximation adopted in Ref.~\cite{Smith:2011ev}.
The typical scale of the smoothing can be given by $l_v(z)=v(z)/H(z)$, where $v(z)$ and
$H(z)$ are the thermal velocity of WDM and the Hubble rate, respectively.
According to Ref.~\cite{Smith:2011ev}, $v(z)$ can be approximated as 
\begin{equation}
v(z)=0.036 \,{\rm km/s} (1+z)\,
\left(\frac{\Omega_{\rm WDM}h^2}{0.11}\right)^{1/3}
\left(\frac{m_{\rm WDM}}{\rm keV}\right)^{-4/3}, 
\end{equation}
which leads to $l_v(z)=1.1\, {\rm kpc}\, (1+z)^{1/2}\left(
\frac{\Omega_{\rm WDM}h^2}{0.11}\right)^{-1/6}
\left(\frac{m_{\rm WDM}}{\rm keV}\right)^{-4/3}$
in the $\Lambda$WDM model.
Within this length, density profile may be smoothed.
Following Ref. \cite{Smith:2011ev}, to obtain smoothed density profiles, 
we convolved the TIS profile with 
a Gaussian kernels with width $l_v(z)$.

\begin{figure}
  \begin{center}
    \begin{tabular}{cc}
    \hspace{-5mm}\scalebox{.75}{\includegraphics{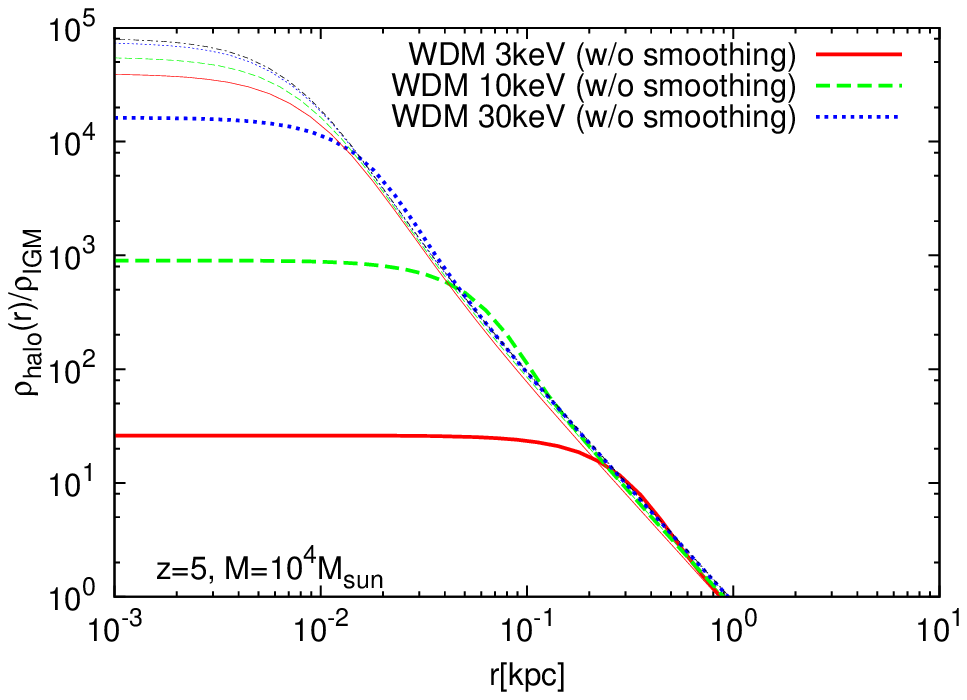}} &
    \hspace{-5mm}\scalebox{.75}{\includegraphics{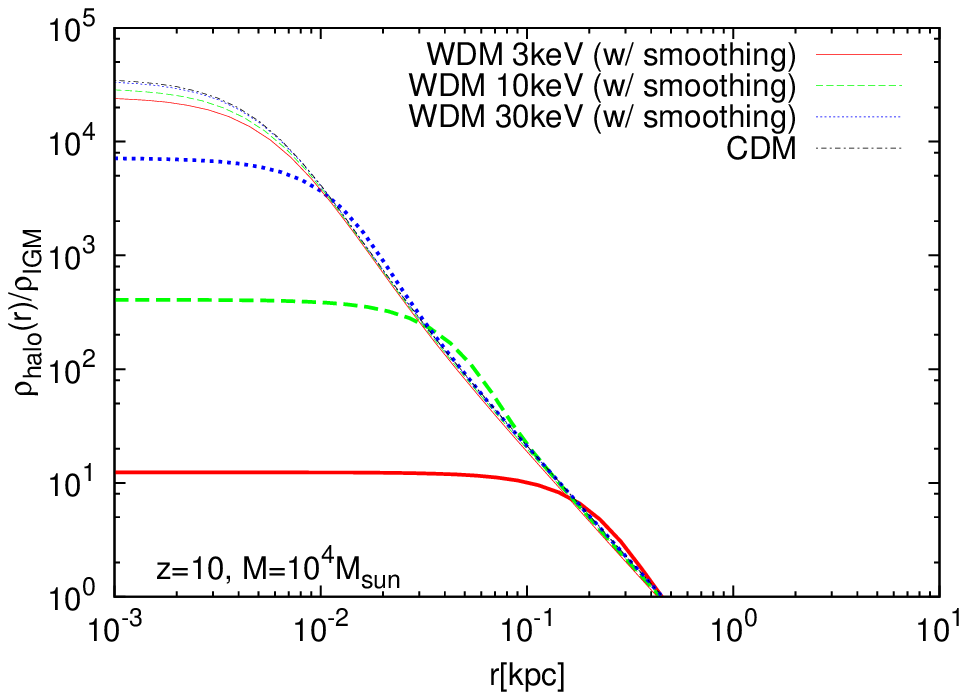}} \\
    \hspace{-5mm}\scalebox{.75}{\includegraphics{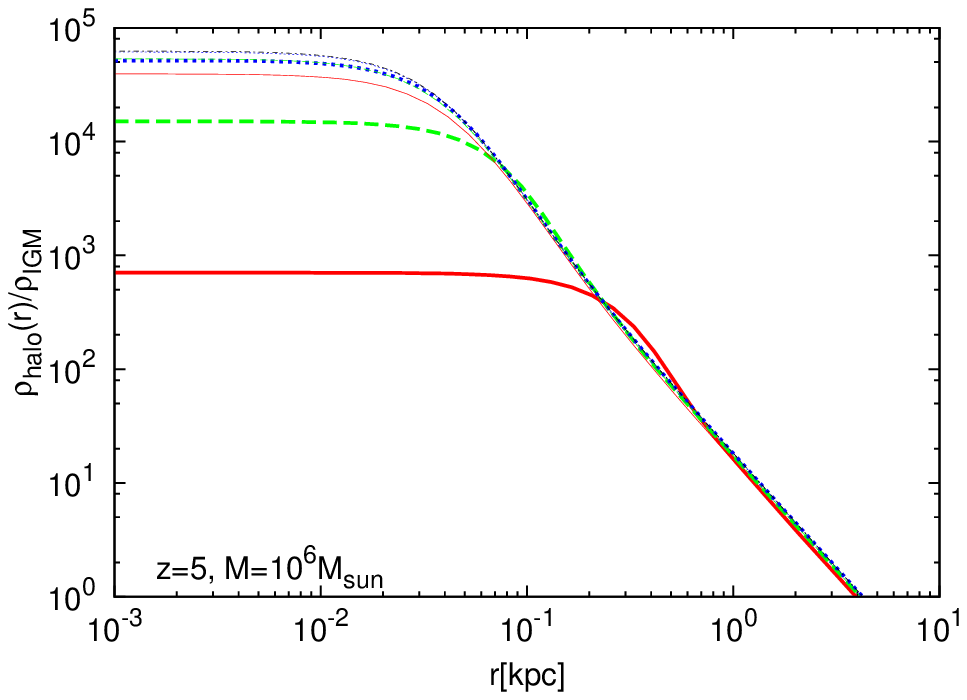}} &
    \hspace{-5mm}\scalebox{.75}{\includegraphics{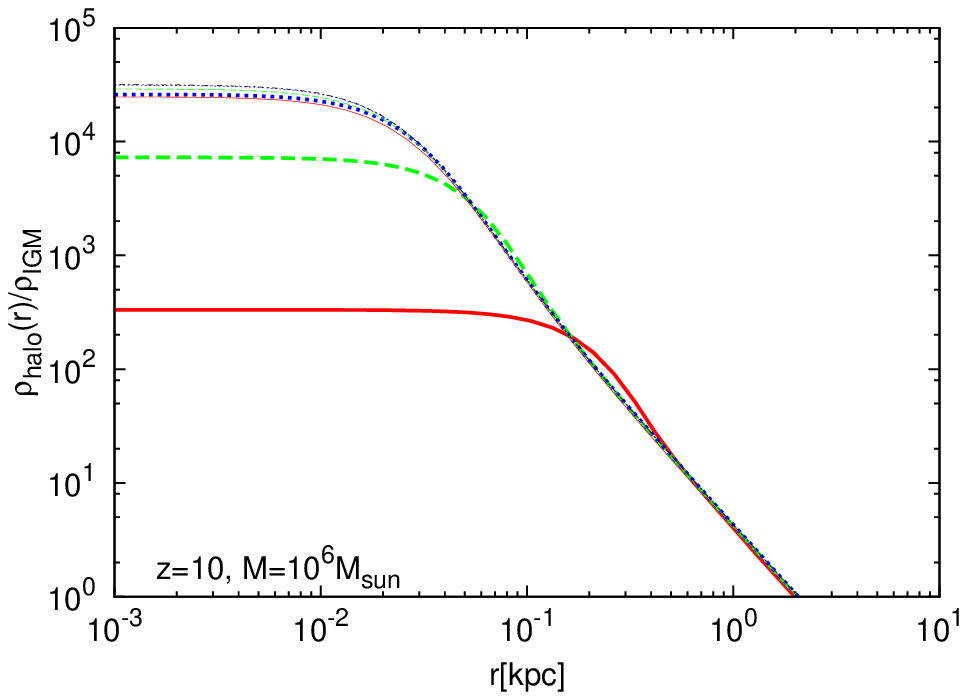}} \\
    \hspace{-5mm}\scalebox{.75}{\includegraphics{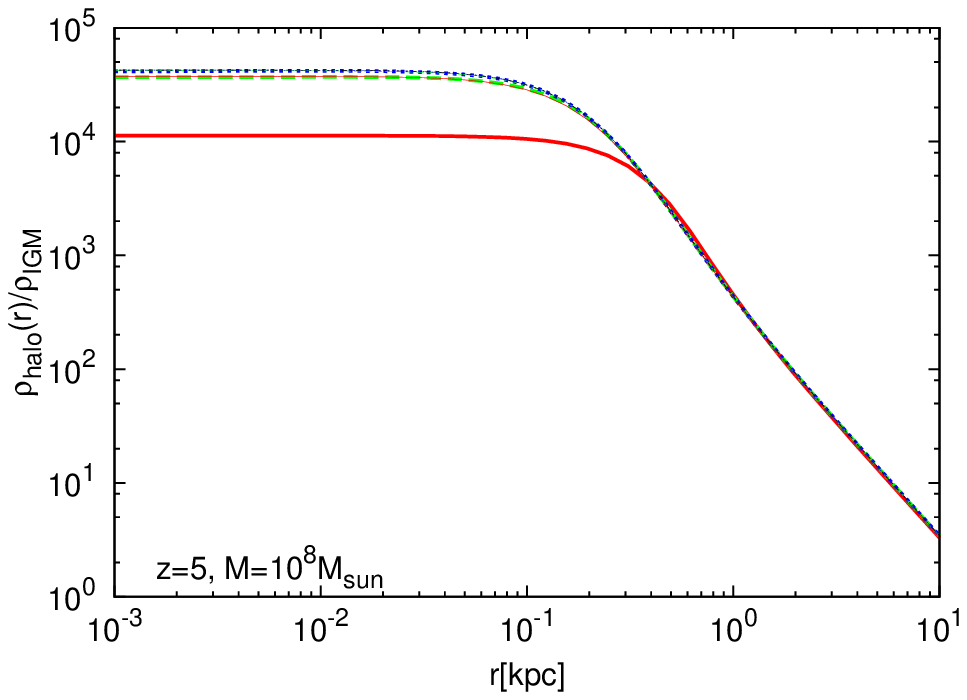}} &
    \hspace{-5mm}\scalebox{.75}{\includegraphics{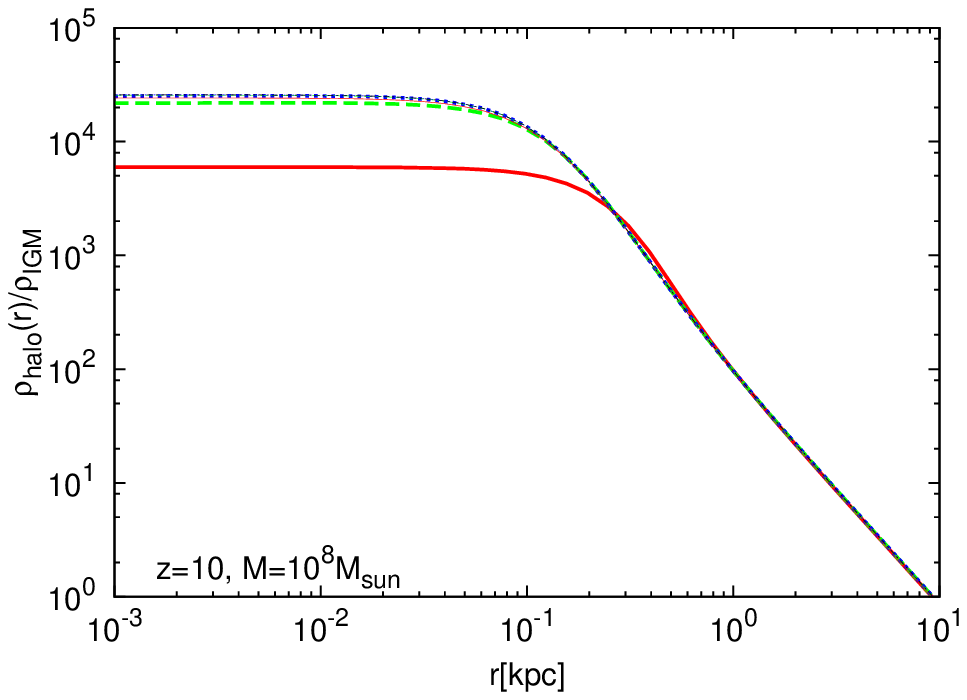}} 
    \end{tabular}
  \end{center}
  \caption{Comparison of the gas density profile between WDM and CDM models.
  Shown are the gas density normalized by that of IGM as function of distance from the halo center.
  Cases of WDM with $m_{\rm WDM}=3$~keV (red), $10$~keV (green) $30$~keV (blue)
  are plotted as well as CDM one (black thin line).
  In order from top to bottom, the mass of haloes are set to $M=10^4$, $10^6$ and $10^8M_\odot$, respectively.
  Left and right panels respectively correspond to redshifts $z=5$ and $z=10$.  
  In the cases of WDM models, the profiles with and without the smoothing due to thermal velocity are depicted
  in thick and thin lines, respectively.
  Note that, as the WDM mass $m_{\rm WDM}$ or the halo mass $M$ becomes larger, 
  the profile becomes closer to that in the CDM model.
  Moreover, without smoothing due to thermal velocity, 
  profiles in WDM models deviate little from that in the CDM one, 
  which causes that thin lines in the figures are difficult to be distinguished from one another.
  }
  \label{fig:profile}
\end{figure}

In Fig.~\ref{fig:profile}, we compare halo profiles in WDM models
with that in the CDM one. 
In the figure, gas density in a halo $\rho_{\rm halo}$ divided by the mean baryon density $\rho_{\rm IGM}$
is plotted as function of distance from the halo center $r$ for several redshifts and halo masses.
From the figure, we can see that the smaller mass of a halo is,
the more prominent the WDM effect on the halo profile is.
In particular, for $m_{\rm WDM}=3$~keV, the maximum density is 
three orders of magnitude smaller than in CDM model, while, as for the size of
halos, it is an order of magnitude larger than in CDM.

Let us see  the two effects on 21~cm line fluctuations
which we discussed above: 
(A) presence of cut-off in the halo mass function at $M_{\rm fs}$
and (B) smoothing in profile due to thermal velocity.
In the following, we examine three cases (I $\sim$ III) as variants from the baseline one.
In table Table~\ref{tbl:variants}, we summarize which effects are included or not in each case.
In Cases I and II , we respectively add the effects (A) and (B) to the baseline one.
In Case III, where we added both the effects (A) and (B) to the baseline one, the effects of
WDM are maximally taken into account.

\begin{table}
  \begin{center}
    \begin{tabular}{lcccc}
      \hline\hline
      & baseline & I & II & III \\
      \hline
      (A) mass cut-off & & \checkmark & & \checkmark \\
      (B) smoothed profile & & & \checkmark & \checkmark \\
      \hline\hline
    \end{tabular}
  \end{center}
  \caption{Correspondences between variant cases examined in Section~\ref{sec:profile} 
  and the effects included in each of them.
  }
  \label{tbl:variants}
\end{table}

\begin{figure}
  \begin{center}
    \hspace{-5mm}\scalebox{.8}{\includegraphics{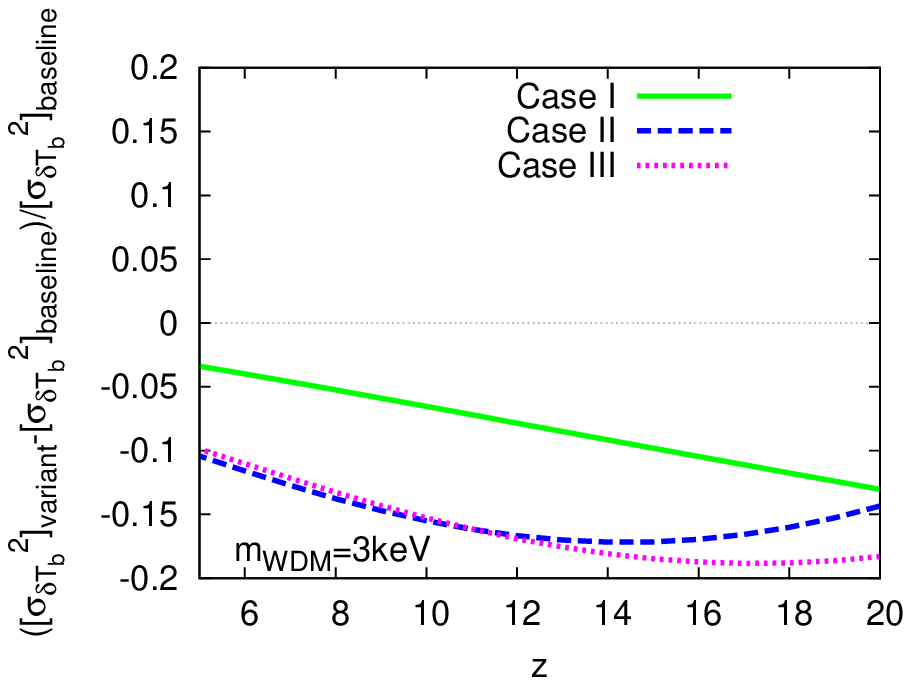}} \\
    \hspace{-5mm}\scalebox{.8}{\includegraphics{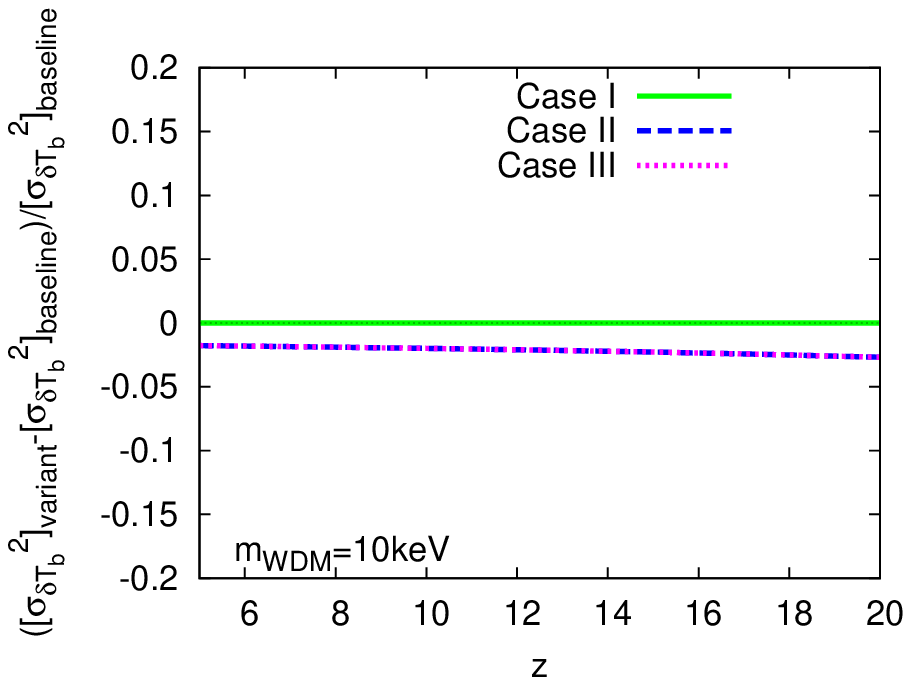}} \\
    \hspace{-5mm}\scalebox{.8}{\includegraphics{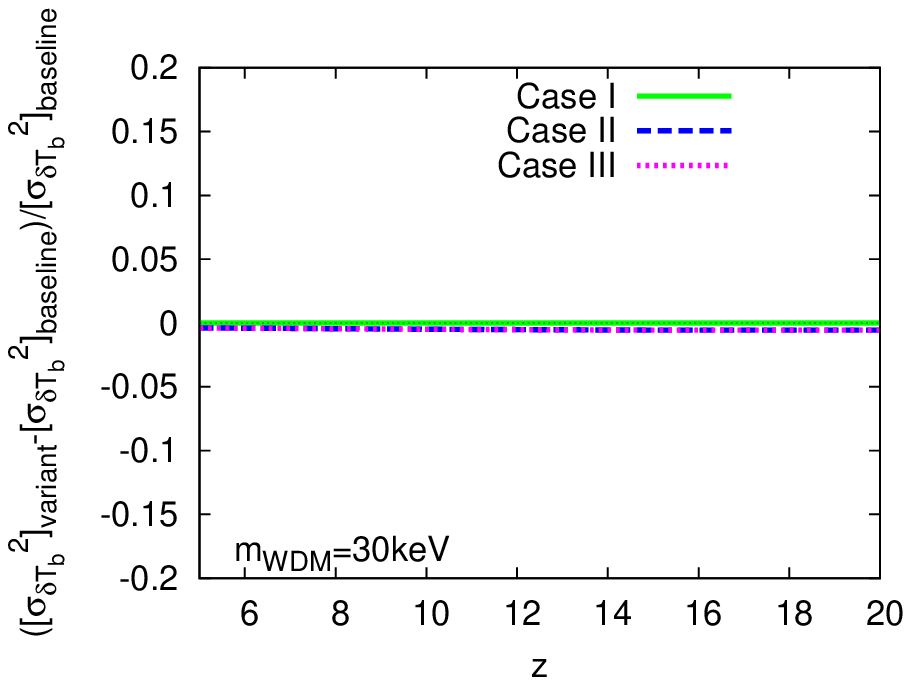}}
  \end{center}
  \caption{Impact of WDM effects on formation and profile of minihalos.
  Shown are the fractional difference in $\sigma_{\delta T_b}^2$ 
  between variants and baseline models. From top to bottom, cases of $m_{\rm WDM}=3$, 10, 30~keV
  are plotted and in each panel, Cases I, II and III are depicted in green solid, blue dashed and magenta dotted lines, respectively.
  Note that in the middle and bottom panels, the green solid line (Case I) is almost identical with the $x$-axis (baseline), 
  and the blue dashed (Case II) and magenta dotted (Case III) lines are also indistinguishable.
  }
  \label{fig:diff}
\end{figure}

In Fig.~\ref{fig:diff}, we plot fractional differences in $\sigma_{\delta T_b}^2$
for three variant cases against the baseline one.
It is clearly seen that the fractional differences tend to be larger as $m_{\rm WDM}$ becomes small.
However, even if we take $m_{\rm WDM}=3$~keV, which
roughly corresponds to the current lower limits \cite{Seljak:2006qw,Viel:2007mv,Viel:2013fqw}, 
these effects change $\sigma_{\delta T_b}^2$ at most a few tens of percents.
On the other hand, if we consider $m_{\rm WDM}=30$~keV, which is roughly the lower bounds
expected for SKA and FFTT, the fractional differences are merely of a few percents.
As far as we are concerning 21~cm observations whose sensitivities are
comparable or better than SKA, detailed effects of WDM on formation and structures of halos are
not relevant.

As a closing remark of this section, let us comment about 
precise prediction on the density profiles of small halos and halo temperature.
Note that,  roughly speaking, halos whose profiles can be drastically changed
by WDM have masses comparable or smaller than $M_{\rm fs}$.
By comparing Cases II and III, we can see contributions of such 
small halos with smoothed profiles. The difference between
these two cases can be recognized only for small $m_{\rm WDM}\lesssim 3$\, keV
and at high redshifts ($z\gtrsim10$), which is difficult to be observed~(see the sensitivity levels
of SKA and FFTT in Fig.~\ref{fig:rms}).
At lower redshifts, or for larger $m_{\rm WDM}$, we can hardly
see the difference. Therefore we conclude that precise predictions 
on smoothed profile of small halos are not important in rms of 21~cm line fluctuations.
We however note that if we want to probe $m_{\rm WDM}\lesssim 3$~keV, 
precise predictions would be important.

As to the halo temperature, while we only consider the modification of density profiles due to WDM,
the temperature is fixed at the one in a ``TIS'' model of CDM. Because
the temperature depends on the matter density profile,
the smoothing of density profiles also change the halo temperature.
In order to obtain the temperature appropriate to the smoothing,
it is required to solve the hydrostatic condition with smoothed matter
density profile, but this is beyond the scope of this paper.
However, since the modification on 21~cm fluctuations due to the
smoothed density profile
is negligible for mass range of our interest ($m_{\rm WDM} >10$~keV), 
the temperature effect could be small.

\section{Summary and discussions}
\label{sec:summary}

In this paper, we have investigated 21~cm line fluctuations 
from minihalos in WDM models. The linear matter power spectrum in a WDM model
is suppressed on small scales due to the free-streaming of WDM.
Therefore, we have found that 
WDM with $m_{\rm WDM}\simeq10$~keV 
significantly reduces abundance of minihalos and delay 
their formation.
Having seen this, we have computed 21~cm line fluctuations from 
minihalos and shown that, compared with the CDM model,  they are suppressed in WDM models.
We also have investigated the feasibility to constrain the mass of WDM
particle by future 21~cm surveys.
We have found that
the deviation from the CDM model can be observed by future 
21~cm surveys, and SKA and FFTT can respectively give lower bounds $m_{\rm WDM}>24$~keV and 31~keV, if
the minihalo signal can be observable above $z=5$.
On the other hand, for larger mass, it would be difficult to distinguish WDM models 
from the CDM one using the rms of 21~cm line fluctuations from minihalos.
Interesting WDM mass range to solve so-called small-scale problems in the
structure formations is 
$m_{\rm WDM}=\mathcal O(1)$~keV.
Therefore, WDM models might be ruled out 
as a solution of small-scale problems by future 21~cm surveys.

We have also examined other effects which are unique to WDM models, such
as the presence of a cut-off in a mass function at the free-streaming
mass and smoothing of halo profiles due to the non-negligible thermal
velocity of WDM. However, we have found that the modification due to these effects on 21~cm fluctuations are small.
In particular, for $m_{\rm WDM}>10$~keV, these effects are negligible.

In our analysis, we omitted masses of active neutrinos, whose effects can in principle 
degenerate with those of WDM as both of them affect structure formation through free-streaming.
However, as far as one considers masses of neutrinos allowed by current 
observational constraints, say, $m_\nu\lesssim$0.3~eV\footnote{For recent review, we refer to e.g. 
Ref.~\cite{Lesgourgues:2012uu}}, their effects on 
21~cm line emission are not very significant.
This is because energy density of massive neutrinos
in the late-time universe is at most a few tenth of percents of that of DM and suppression on structure formation within their
free-streaming scales are much less prominent than of WDM; they do not create a cut-off, 
but only a small step-like feature in the matter transfer function. Lacking significant suppression on
matter fluctuations on scales of minihalos, massive active neutrinos cannot significantly affect
21~cm line fluctuations from minihalos.
Therefore we conclude that there should not be 
a significant degeneracy between masses of WDM and active neutrinos in a
pure WDM model.

Regarding the choice of a prescription for mass function, 
we have adopted the Sheth-Tormen mass function throughout this paper,
while
precise mass function of minihalos is still under debate.
Refs.~\cite{Chongchitnan:2012we,Takeuchi:2013hza}
have studied the effect of mass function on 21~cm fluctuations due to minihalos,
although they have done in context other than WDM.
They have shown that
the Sheth-Tormen mass function tends to give moderate signals of 21~cm line fluctuations from minihalos.
Therefore, adopting other mass functions, such as the extended Press-Schechter one, would give 
larger signals, which leads to tighter constraints on $m_{\rm WDM}$.

We would also like to comment on dependence of halo profiles. 
Here we adopted the TIS profile, which is smooth at centers of halos
even in the CDM model. On the other hand, if one adopt other profiles which are steeper at halo centers, 
e.g.,~the NFW profile~\cite{Navarro:1995iw}, the smoothing effects of WDM would appear more 
prominent. To this respect, our analysis may have underestimated the improvement
of the bounds on $m_{\rm WDM}$ due to the smoothing effects on profile
of WDM.
We save this issue for future studies.

Our analysis is fully based on analytical computation and the results presented in this paper
may be more or less modified by more detailed studies based on numerical simulations.
For instance, in Ref.~\cite{Boehm:2003xr} it is suggested that the regeneration mechanism due to the nonlinear evolution of matter perturbations can
blunt the suppression in the matter power spectrum due to the free-streaming of WDM to greater extent at lower redshifts.
However, the abundance of halos and hence the 21~cm rms from minihalos may be more tolerant than the matter power spectrum 
to the regeneration mechanism~\cite{Boehm:2003xr}.
Although more detailed analyses are necessary prior to derive constraints from actual observations, 
our results demonstrate a remarkable potential of 21~cm line fluctuations 
from minihalos in probing unknown nature of dark matter.

\acknowledgments
TS is supported by the Academy of Finland grant 1263714.
HT is supported by the DOE at ASU.

\providecommand{\href}[2]{#2}\begingroup\raggedright


\begin{thebibliography}{10}

\bibitem{Frenk:2012ph} 
  C.~S.~Frenk and S.~D.~M.~White,
  Annalen Phys.\  {\bf 524}, 507 (2012)
  [arXiv:1210.0544 [astro-ph.CO]].

\bibitem{Weinberg:2013aya} 
  D.~H.~Weinberg, J.~S.~Bullock, F.~Governato, R.~K.~de Naray and A.~H.~G.~Peter,
  arXiv:1306.0913 [astro-ph.CO].
	
\bibitem{Klypin:1999uc} 
  A.~A.~Klypin, A.~V.~Kravtsov, O.~Valenzuela and F.~Prada,
  Astrophys.\ J.\  {\bf 522}, 82 (1999)
  [astro-ph/9901240].
\bibitem{Moore:1999nt} 
  B.~Moore, S.~Ghigna, F.~Governato, G.~Lake, T.~R.~Quinn, J.~Stadel and P.~Tozzi,
  Astrophys.\ J.\  {\bf 524}, L19 (1999)
  [astro-ph/9907411].

\bibitem{Navarro:1996gj} 
  J.~F.~Navarro, C.~S.~Frenk and S.~D.~M.~White,
  Astrophys.\ J.\  {\bf 490}, 493 (1997)
  [astro-ph/9611107].

\bibitem{Moore:1999gc} 
  B.~Moore, T.~R.~Quinn, F.~Governato, J.~Stadel and G.~Lake,
  Mon.\ Not.\ Roy.\ Astron.\ Soc.\  {\bf 310}, 1147 (1999)
  [astro-ph/9903164].

	
\bibitem{Flores:1994gz} 
  R.~A.~Flores and J.~R.~Primack,
  Astrophys.\ J.\  {\bf 427}, L1 (1994)
  [astro-ph/9402004].

\bibitem{Moore:1994yx} 
  B.~Moore,
  Nature {\bf 370}, 629 (1994).
	

\bibitem{SommerLarsen:1999jx} 
  J.~Sommer-Larsen and A.~Dolgov,
  Astrophys.\ J.\  {\bf 551}, 608 (2001)
  [astro-ph/9912166].

\bibitem{Hogan:2000bv} 
  C.~J.~Hogan and J.~J.~Dalcanton,
  Phys.\ Rev.\ D {\bf 62}, 063511 (2000)
  [astro-ph/0002330].

\bibitem{Schaeffer:1988}
  R.~Schaeffer and J.~Silk, 
  Astrophys.\ J.\  {\bf 332}, 1 (1988)
  [astro-ph/9508025].
  
\bibitem{White:2000sy} 
  M.~J.~White and R.~A.~C.~Croft,
  Astrophys.\ J.\  {\bf 539}, 497 (2000)
  [astro-ph/0001247].

\bibitem{Colin:2000dn} 
  P.~Colin, V.~Avila-Reese and O.~Valenzuela,
  Astrophys.\ J.\  {\bf 542}, 622 (2000)
  [astro-ph/0004115].

\bibitem{Colin:2007bk} 
  P.~Colin, O.~Valenzuela and V.~Avila-Reese,
  Astrophys.\ J.\  {\bf 673}, 203 (2008)
  [arXiv:0709.4027 [astro-ph]].
	
\bibitem{Narayanan:2000tp} 
  V.~K.~Narayanan, D.~N.~Spergel, R.~Dave and C.~-P.~Ma,
  astro-ph/0005095.

\bibitem{Seljak:2006qw} 
  U.~Seljak, A.~Makarov, P.~McDonald and H.~Trac,
  Phys.\ Rev.\ Lett.\  {\bf 97}, 191303 (2006)
  [astro-ph/0602430].
  
\bibitem{Viel:2007mv} 
  M.~Viel, G.~D.~Becker, J.~S.~Bolton, M.~G.~Haehnelt, M.~Rauch and W.~L.~W.~Sargent,
  Phys.\ Rev.\ Lett.\  {\bf 100}, 041304 (2008)
  [arXiv:0709.0131 [astro-ph]].

\bibitem{Viel:2013fqw} 
  M.~Viel, G.~D.~Becker, J.~S.~Bolton and M.~G.~Haehnelt,
  Physical Review D {\bf 88}, no. 4, 043502 (2013)
  [arXiv:1306.2314 [astro-ph.CO]].


\bibitem{deSouza:2013wsa} 
  R.~S.~de Souza, A.~Mesinger, A.~Ferrara, Z.~Haiman, R.~Perna and N.~Yoshida,
  MNRAS, 432, {\bf 3218} (2013)
  [arXiv:1303.5060 [astro-ph.CO]].
	
	
\bibitem{Markovic:2010te} 
  K.~Markovic, S.~Bridle, A.~Slosar and J.~Weller,
  JCAP {\bf 1101}, 022 (2011)
  [arXiv:1009.0218 [astro-ph.CO]].

\bibitem{Smith:2011ev} 
  R.~E.~Smith and K.~Markovic,
  Phys.\ Rev.\ D {\bf 84}, 063507 (2011)
  [arXiv:1103.2134 [astro-ph.CO]].


\bibitem{Yoshida:2003rm} 
  N.~Yoshida, A.~Sokasian, L.~Hernquist and V.~Springel,
  Astrophys.\ J.\  {\bf 591}, L1 (2003)
  [astro-ph/0303622].

\bibitem{Yue:2012na} 
  B.~Yue and X.~Chen,
  Astrophys.\ J.\  {\bf 747}, 127 (2012)
  [arXiv:1201.3686 [astro-ph.CO]].

	
\bibitem{Boyarsky:2007ay} 
  A.~Boyarsky, D.~Iakubovskyi, O.~Ruchayskiy and V.~Savchenko,
  Mon.\ Not.\ Roy.\ Astron.\ Soc.\  {\bf 387}, 1361 (2008)
  [arXiv:0709.2301 [astro-ph]].

	
\bibitem{Boyarsky:2008mt} 
  A.~Boyarsky, J.~Lesgourgues, O.~Ruchayskiy and M.~Viel,
  Phys.\ Rev.\ Lett.\  {\bf 102}, 201304 (2009)
  [arXiv:0812.3256 [hep-ph]].


\bibitem{Loeb:2003ya} 
  A.~Loeb and M.~Zaldarriaga,
  Phys.\ Rev.\ Lett.\  {\bf 92}, 211301 (2004)
  [astro-ph/0312134].
  
\bibitem{Sitwell:2013fpa} 
  M.~Sitwell, A.~Mesinger, Y.~-Z.~Ma and K.~Sigurdson,
  arXiv:1310.0029 [astro-ph.CO].

\bibitem{Iliev:2002gj} 
  I.~T.~Iliev, P.~R.~Shapiro, A.~Ferrara and H.~Martel,
  Astrophys.\ J.\  {\bf 572}, 123 (2002)
  [astro-ph/0202410].

\bibitem{Furlanetto:2002ng} 
  S.~Furlanetto and A.~Loeb,
  Astrophys.\ J.\  {\bf 579}, 1 (2002)
  [astro-ph/0206308].

	
\bibitem{Tegmark:2008au} 
  M.~Tegmark and M.~Zaldarriaga,
  Phys.\ Rev.\ D {\bf 79}, 083530 (2009)
  [arXiv:0805.4414 [astro-ph]].

\bibitem{Takeuchi:2013hza} 
  Y.~Takeuchi and S.~Chongchitnan,
  arXiv:1311.2585 [astro-ph.CO].
	
\bibitem{Sekiguchi:2013lma} 
  T.~Sekiguchi, H.~Tashiro, J.~Silk and N.~Sugiyama,
  arXiv:1311.3294 [astro-ph.CO].

\bibitem{Chongchitnan:2012we} 
  S.~Chongchitnan and J.~Silk,
  arXiv:1205.6799 [astro-ph.CO].
	
\bibitem{Tashiro:2013xra} 
  H.~Tashiro, T.~Sekiguchi and J.~Silk,
  arXiv:1310.4176 [astro-ph.CO].

\bibitem{Bode:2000gq} 
  P.~Bode, J.~P.~Ostriker and N.~Turok,
  Astrophys.\ J.\  {\bf 556}, 93 (2001)
  [astro-ph/0010389].

\bibitem{Boyanovsky:2010pw} 
  D.~Boyanovsky and J.~Wu,
  Phys.\ Rev.\ D {\bf 83}, 043524 (2011)
  [arXiv:1008.0992 [astro-ph.CO]].

\bibitem{Lewis:1999bs} 
  A.~Lewis, A.~Challinor and A.~Lasenby,
  Astrophys.\ J.\  {\bf 538}, 473 (2000)
  [astro-ph/9911177].

\bibitem{Howlett:2012mh} 
  C.~Howlett, A.~Lewis, A.~Hall and A.~Challinor,
  JCAP {\bf 1204}, 027 (2012)
  [arXiv:1201.3654 [astro-ph.CO]].

\bibitem{Press:1973iz} 
  W.~H.~Press and P.~Schechter,
  Astrophys.\ J.\  {\bf 187}, 425 (1974).
  
\bibitem{Sheth:1999mn} 
  R.~K.~Sheth and G.~Tormen,
  Mon.\ Not.\ Roy.\ Astron.\ Soc.\  {\bf 308}, 119 (1999)
  [astro-ph/9901122].

\bibitem{Shapiro:1998zp} 
  P.~R.~Shapiro and I.~T.~Iliev,
  Mon.\ Not.\ Roy.\ Astron.\ Soc.\  {\bf 307}, 203 (1999)
  [astro-ph/9810164].

\bibitem{Dodelson:2003ft} 
  S.~Dodelson,
  Amsterdam, Netherlands: Academic Pr. (2003) 440 p
	
\bibitem{Furlanetto:2006jb} 
  S.~Furlanetto, S.~P.~Oh and F.~Briggs,
  Phys.\ Rept.\  {\bf 433}, 181 (2006)
  [astro-ph/0608032].
	
\bibitem{Zavala:2009ms} 
  J.~Zavala, Y.~P.~Jing, A.~Faltenbacher, G.~Yepes, Y.~Hoffman, S.~Gottlober and B.~Catinella,
  Astrophys.\ J.\  {\bf 700}, 1779 (2009)
  [arXiv:0906.0585 [astro-ph.CO]].

\bibitem{Polisensky:2010rw} 
  E.~Polisensky and M.~Ricotti,
  Phys.\ Rev.\ D {\bf 83}, 043506 (2011)
  [arXiv:1004.1459 [astro-ph.CO]].

\bibitem{Benson:2012su} 
  A.~J.~Benson, A.~Farahi, S.~Cole, L.~A.~Moustakas, A.~Jenkins, M.~Lovell, R.~Kennedy and J.~Helly {\it et al.},
  arXiv:1209.3018 [astro-ph.CO].

\bibitem{Schneider:2013ria} 
  A.~Schneider, R.~E.~Smith and D.~Reed,
  arXiv:1303.0839 [astro-ph.CO].

\bibitem{Angulo:2013sza} 
  R.~E.~Angulo, O.~Hahn and T.~Abel,
  arXiv:1304.2406 [astro-ph.CO].

\bibitem{Wang:2007he} 
  J.~Wang and S.~D.~M.~White,
  Mon.\ Not.\ Roy.\ Astron.\ Soc.\  {\bf 380}, 93 (2007)
  [astro-ph/0702575 [ASTRO-PH]].

\bibitem{Lesgourgues:2012uu} 
  J.~Lesgourgues and S.~Pastor,
  Adv.\ High Energy Phys.\  {\bf 2012}, 608515 (2012)
  [arXiv:1212.6154 [hep-ph]].

\bibitem{Navarro:1995iw} 
  J.~F.~Navarro, C.~S.~Frenk and S.~D.~M.~White,
  Astrophys.\ J.\  {\bf 462}, 563 (1996)
  [astro-ph/9508025].

\bibitem{Boehm:2003xr} 
  C.~Boehm, H.~Mathis, J.~Devriendt and J.~Silk,
  Mon.\ Not.\ Roy.\ Astron.\ Soc.\  {\bf 360(1)}, 282 (2005)
  [astro-ph/0309652].

\end{thebibliography}
\end{document}